\documentclass[trackchanges,twocolumn]{aastex701}

\begin{document}

\title{An extra hard spectral component peaking at sub-GeV  in the prompt emission of GRB 260226A}

\author[0000-0002-6036-985X]{Yi-Yun Huang}
\affiliation{School of Astronomy and Space Science, Nanjing University, Nanjing 210023, China}
\affiliation{Key laboratory of Modern Astronomy and Astrophysics (Nanjing University), Ministry of Education, Nanjing 210023, China}
\email{None}

\author[0000-0001-6863-5369]{Hai-Ming Zhang}
\affiliation{Guangxi Key Laboratory for Relativistic Astrophysics, School of Physical Science and Technology, Guangxi University, 530004 Nanning, Guangxi, China}
\email{None}

\author[0000-0003-1576-0961]{Ruo-Yu Liu}
\affiliation{School of Astronomy and Space Science, Nanjing University, Nanjing 210023, China}
\affiliation{Key laboratory of Modern Astronomy and Astrophysics (Nanjing University), Ministry of Education, Nanjing 210023, China}
\affiliation{Tianfu Cosmic Ray Research Center, Chengdu 610000,Sichuan, China}
\email{None}

\author[0000-0002-5881-335X]{Xiang-Yu Wang}
\affiliation{School of Astronomy and Space Science, Nanjing University, Nanjing 210023, China}
\affiliation{Key laboratory of Modern Astronomy and Astrophysics (Nanjing University), Ministry of Education, Nanjing 210023, China}
\affiliation{Tianfu Cosmic Ray Research Center, Chengdu 610000,Sichuan, China}
\email[show]{xywang@nju.edu.cn}

\begin{abstract}
The prompt emission spectra of gamma-ray bursts (GRBs) have long been empirically described by the Band function over the keV--MeV range, whereas several \textit{Fermi}/LAT-detected GRBs show evidence for an extra hard component at higher energies. 
Here we present a  joint GBM--LAT study of GRB~260226A,  a rare LAT
seeded onboard trigger GRB, and find that an extra sub-GeV component is present.  In the time-resolved analysis, we find that  a cutoff power-law  model fits the spectral data of the extra component better than other models at earlier times, while at later times, the broken  power-law model is preferred (or at least equally good).  Interpreting the early cutoff as the $\gamma\gamma$ absorption implies that the Lorentz factor at early time is lower and increases with time.
The low ratio between the peak energy of the extra component and that of the Band component is difficult to explain with a one-zone synchrotron self-Compton (SSC) scenario. Two-zone emission models, such as external inverse Compton scattering of the photosphere emission by relativistic electrons accelerated in  internal shocks, could provide a possible explanation. 
\end{abstract}


\section{Introduction} 

Gamma-ray bursts (GRBs) are among the most luminous electromagnetic transients in the Universe, releasing up to $\sim 10^{54}$~erg on timescales ranging from milliseconds to thousands of seconds. 
The spectra during their prompt emission in the keV--MeV range are commonly described by the empirical Band function \citep{Band1993}.
While it is generally accepted that the prompt emission results from internal dissipation within the relativistic jets, such as internal shocks \citep{Rees1994} or magnetic reconnection processes \citep{Zhang2011}, the origin of the Band function remains unknown. Synchrotron emission from relativistic electrons has been widely discussed as a possible origin of the Band spectrum \citep[e.g.,][]{Tavani1996,Lloyd2000,Uhm2014,Burgess2020}, while Comptonized photospheric emission provides another viable explanation \citep[e.g.,][]{MeszarosRees2000,PeEr2006,Beloborodov2010,Ryde2010}.

With the launch of the \textit{Fermi} Gamma-ray Space Telescope \citep{Fermi2009}, especially the Large Area Telescope (LAT; \citealt{Atwood2009}), the observable spectral window has been extended to GeV energies. 
Joint spectroscopy with the \textit{Fermi} Gamma-ray Burst Monitor (GBM) and LAT has revealed that a non-negligible fraction of LAT-detected bursts cannot be described by a single Band component \citep{GRBcatalog19}.
Instead, they require an additional hard spectral component at high energies.
Clear examples include GRB 090902B \citep{Abdo2009b}, GRB 090510 \citep{Ackermann2010}, GRB 090926A \citep{Ackermann2011}, GRB 110721A \citep{Ackermann2013b}, GRB 130427A \citep{Ackermann2014}, GRB 131108A \citep{Chen2014}, GRB 190114C \citep{Ajello2020,Chand2020}, and GRB 240825A \citep{Zhang2025a}. Among them, three GRBs (i.e., GRB~090926A, GRB~190114C, GRB~240825A) show a spectral peak or break in the extra components, while others show a simple power-law extending to the high-energy end without any observed break. 
GRB 090926A was the first GRB with a significant cutoff/break detected in the extra hard component, with a cutoff energy of $\sim1.4$ GeV \citep{Ackermann2011}. This cutoff has been interpreted as possible internal $\gamma\gamma$ opacity, allowing an estimate of the bulk Lorentz factor of the emitting region \citep{Ackermann2011,Yassine2017}.

Recently, \citet{Zhang2025a} found that GRB 240825A has a two-hump prompt spectrum, composed of a Band component and an extra hard component peaking at sub-GeV energies ($\sim$40~MeV). The extra component was better described by a broken power law (BPL) than by a cutoff power-law (CPL), favoring an intrinsic spectral break. \citet{Ajello2020} found that GRB 190114C has an extra component at earlier times, and the extra component is well described by  CPL, which could be due to $\gamma\gamma$ absorption within the source. 
The cutoff energy is observed to increase with time before disappearing entirely at later times\citep{Ajello2020}.

In this paper, we report on a detailed joint GBM+LAT analysis of GRB 260226A,  a rare LAT-seeded onboard trigger GRB.  LAT-seeded alerts have a very high threshold and this is only the second GRB to have passed this alert after GRB 090510 \citep{Depalo2026GCN43844}.
We show that its prompt emission contains an extra had component peaking at sub-GeV band, making it a new member of the small group of GRBs with extra components peaking at sub-GeV energies. 
In Section \ref{sec:data}, we summarize the observations of GRB 260226A and describe the data reduction.
We show the results of light curves in Section \ref{sec:lc}, as well as the spectral results in Section \ref{sec:sed}. 
We discuss the physical implications in Section~\ref{sec:discussion} and summarize our conclusions in Section~\ref{sec:conclusion}.



\section{Observations and Data Reduction}
\label{sec:data}
\subsection{Observations}

At 10:37:55 UT on 2026 February 26, GBM triggered on and localized the long burst GRB 260226A, with an on-ground position of R.A.~$=45.3^\circ$ and Decl.~$=15.7^\circ$ (J2000), and a statistical uncertainty of $1.0^\circ$ \citep{FermiGBMTeam2026GCN43840}. 
We take the GBM trigger time as $T_0$.
The burst was also independently detected by LAT at 10:38:18.84 UT as a rare LAT-seeded onboard trigger \citep{Depalo2026GCN43844}. 
The refined LAT analysis gave a best on-ground position of R.A.~$=41.93^\circ$ and Decl.~$=7.73^\circ$ (J2000), with a $90\%$ containment radius of $0.11^\circ$. 
During the first 1.5 ks after GBM trigger, the photon flux above 100 MeV was $(1.50\pm0.06)\times10^{-4}\ {\rm ph\ cm^{-2}\ s^{-1}}$, with a photon index of $-3.25\pm0.08$. The highest-energy photon had an energy of 1.1 GeV and arrived at $T_0+420$ s \citep{Depalo2026GCN43850}. 
The GBM light curve shows bright and highly structured prompt emission.
The measured duration is $T_{90}\simeq173$ s in the 50--300 keV band.
The time-averaged GBM spectrum from $T_0$ to $T_0+81$ s is well described by a Band function with $E_{\rm p}=723\pm8$~keV, $\alpha=-0.95\pm0.01$, and $\beta=-2.47\pm0.02$. The 10--1000~keV fluence is $(4.54\pm0.01)\times10^{-4}\ {\rm erg\ cm^{-2}}$, making it one of the brightest GRBs observed by \textit{Fermi}/GBM\citep{Bissaldi2026GCN43851}. GRB 260226A was also detected by AstroSat/CZTI, NuSTAR/SINGS, Glowbug, CALET/CGBM, Insight-HXMT/HE, and Konus-Wind, all of which reported a bright, multi-peaked prompt phase on timescales of tens of seconds \citep{Harsha2026GCN43846,Waratkar2026GCN43854,Woolf2026GCN43855,Kobayashi2026GCN43860,Yu2026GCN43865,Svinkin2026GCN43864}.

\subsection{GBM Data Reduction}

The GBM consists of 12 sodium iodide (NaI)
and two bismuth germanate (BGO) scintillation detectors, covering the energy range from 8 keV to 40 MeV \citep{Meegan2009}. 
For this work, we selected the NaI detectors n0, n1 and n3 and the BGO detector b0,
which had the smallest viewing angles with respect to the GRB position. The NaI data were used in the energy range 8--900 keV, excluding the channels around the iodine K-edge (32--36 keV), while the BGO data were used in the range 0.2--40 MeV. 
We downloaded GBM data from the \textit{Fermi}/GBM public data archive\footnote{https://heasarc.gsfc.nasa.gov/FTP/fermi/data/gbm/daily/}. 
The background was modeled by selecting two off-source intervals before and after the prompt emission, $[-100,-5]$ s and $[600,900]$ s, and fitting them with a polynomial function.

\subsection{LAT Data Reduction and Analysis}

The LAT data of GRB 260226A were downloaded from the Fermi Science Support Center\footnote{https://fermi.gsfc.nasa.gov}. 
We analyzed the Pass 8 data with the Fermi Science Tools (version v2.20) and fermipy (version v1.4.0).

We selected LAT events in the energy range from 100~MeV to 100~GeV within a 15$^\circ$ region of interest centered on the refined LAT position. 
We used the TRANSIENT event class for time bins with durations $\leq200$ s and the SOURCE event class for bins with durations $>200$ s.
To reduce the contamination from the Earth limb, photons with zenith angles larger than 100$^\circ$ were excluded. 
The good-time intervals were selected with the standard filter (DATA\_QUAL$>$0)\&\&(LAT\_CONFIG==1).
The corresponding instrument response functions were P8R3\_TRANSIENT020\_V3 and P8R3\_SOURCE\_V3, respectively.

An unbinned maximum likelihood analysis was performed for the LAT data.
The source model includes GRB 260226A, the Galactic diffuse emission component gll\_iem\_v07.fits, and the isotropic diffuse component appropriate for the selected event class. 
The GRB spectrum was described by a power-law function,
\begin{equation}
    \frac{dN}{dE}=N_0\left(\frac{E}{E_0}\right)^{-\Gamma}.
\end{equation}
The detection significance was estimated with the test statistic ${\rm TS}=2\Delta\ln\mathcal{L}$.

We first analyzed the LAT data in the time interval $T_0$ to $T_0+2000$ s. The best-fit photon index is $\Gamma=3.24 \pm 0.10$, and the corresponding TS value is 2414.46. 
The highest-energy photon associated with GRB 260226A has an energy of 1.1 GeV and arrives at $T_0+433$ s, with an association probability of 99.98\%.

\section{Light Curves}
\label{sec:lc}
\begin{figure*}[htbp]
\centering
\includegraphics[width=\textwidth]{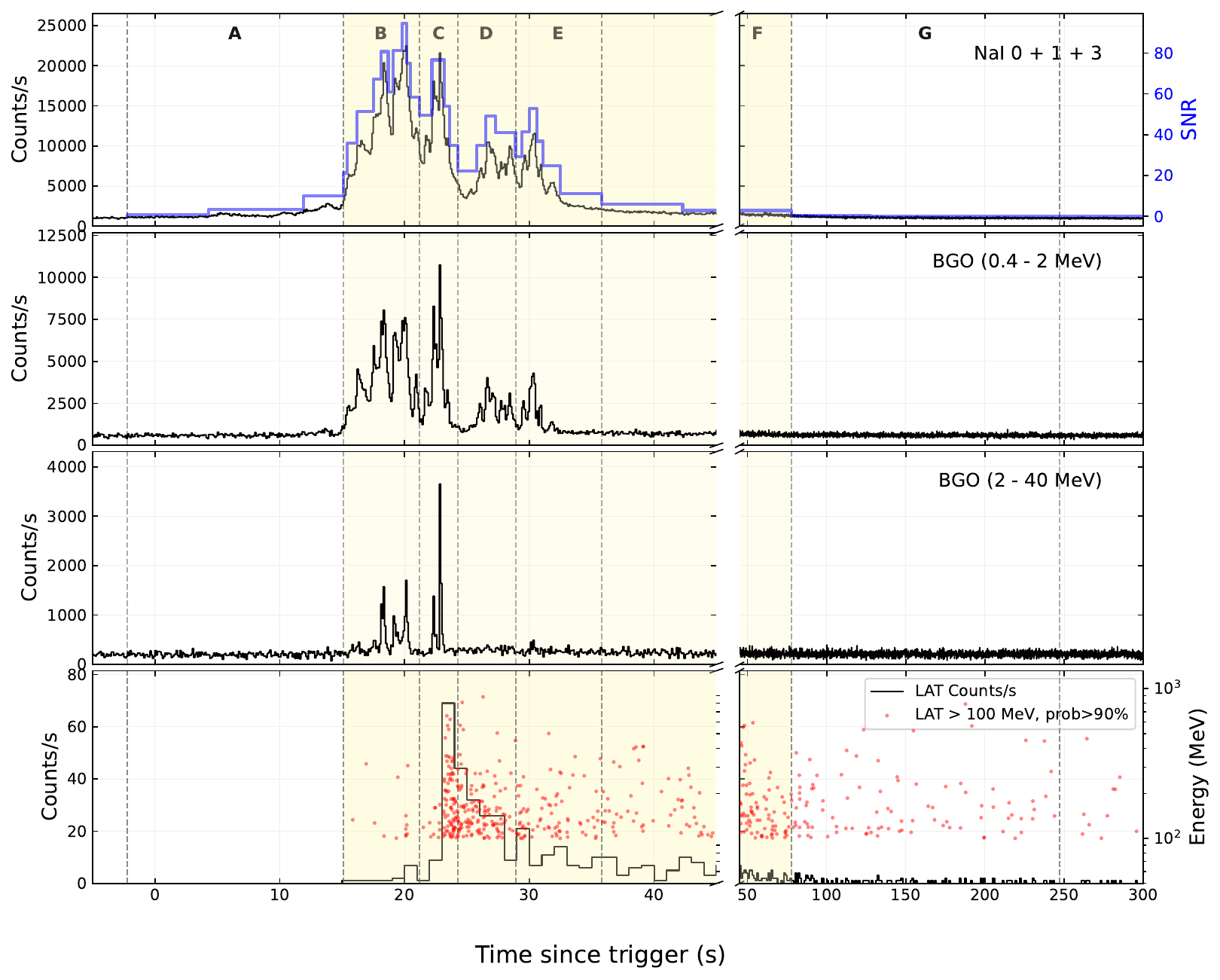}
\caption{GBM and LAT count-rate light curves of GRB 260226A. \textit{Top panel:} GBM light curve in the 8--900~keV energy band. \textit{Middle panels:} GBM light curve in the 200~keV--40~MeV band (BGO). \textit{Bottom panel:} LAT light curve above 100~MeV. Red points mark LAT photons above 100 MeV with probability greater than 90\%. The vertical dashed lines mark the time bins used for time-resolved spectral analysis. The shaded yellow region marks the interval in which the time-resolved spectra
favor a two-component model, as discussed in the section \ref{sec:sed}. }
\label{fig:lc}
\end{figure*}

Figure \ref{fig:lc} presents the GBM and LAT light curves in several energy bands. The GBM light curve is derived from TTE data with a bin size of 0.1 s, while the LAT light curve ($E>100$ MeV) uses a bin size of 1 s.
To determine the time intervals for the spectral analysis, we employ the Bayesian block method \citep{Scargle2013} on the rebinned data of all three NaI detectors. 
The resulting blocks are indicated by the step lines in Figure~\ref{fig:lc}. 
We further group the blocks into seven intervals labeled A--G, with boundaries $T_0+$[-2.2, 15.1, 21.2, 24.3, 28.9, 35.8, 77.6, 247.0] s.
The GBM light curves exhibit multiple sharp pulses during this interval, followed by weaker and more extended emission lasting to hundreds of seconds. 
The LAT emission is detected during the prompt emission phase, however, the peak of GeV emission exhibits a time delay relative to the MeV emission, similar to other LAT-detected GRBs \citep{Ackermann2013b}. 
LAT photons are also detected at later times, extending to $\sim10^3$ s after the trigger when the GBM emission has largely faded. 
The total number of LAT photons with association probability above 90\% is 598 for the period of $T_0$ to $T_0+2000$ s and 303 during the main prompt emission (B--E). 
Only one photon with energy $\geq1$ GeV appears in the very late stage.

\section{Spectral Analysis}
\label{sec:sed}

The joint GBM and LAT spectral fitting is performed with \textsc{XSPEC} v12.15.1 \citep{Arnaud1996}, and C-stat is used as the statistic to estimate uncertainties of the best-fit parameters. 
We perform time-resolved spectral analysis for the time intervals
grouped in section \ref{sec:lc} to study possible spectral evolution.

We test four spectral models for the joint GBM--LAT spectra: a single
Band function, Band+CPL, Band+BPL, and Band plus a smoothly broken
power law (Band+SBPL).
The Band model\citep{Band1993} can be expressed as
\begin{equation}
\resizebox{0.98\columnwidth}{!}{$
\displaystyle
N(E)=
\left\{
\begin{array}{ll}
\displaystyle
A\left(\frac{E}{100\,{\rm keV}}\right)^\alpha
\exp\left(-\frac{E}{E_{\rm 0}}\right),
&
\displaystyle
E<(\alpha-\beta)E_{\rm 0},
\\[2ex]
\displaystyle
A\left[\frac{(\alpha-\beta)E_{\rm 0}}{100\,{\rm keV}}\right]^{\alpha-\beta}
\exp(\beta-\alpha)
\left(\frac{E}{100\,{\rm keV}}\right)^\beta,
&
\displaystyle
E\ge(\alpha-\beta)E_{\rm 0},
\end{array}
\right.
$}
\end{equation}
where $\alpha$ and $\beta$ are low-energy and high-energy photon spectral indices. 
The peak energy $E_{\rm p}$ is related to the e-folding energy $E_{\rm 0}$ through $E_{\rm p}=(2+\alpha)E_{\rm 0}$.

The cutoff power-law model (CPL) is expressed as
\begin{equation}
N(E)=AE^{-\lambda}\exp(-E/E_{\rm c}),
\end{equation}
where $\lambda$ is the power-law photon index below the cutoff energy and $E_{\rm c}$ is the e-folding energy. 

The broken power-law model (BPL) is written as
\begin{equation}
N(E)=
\left\{
\begin{array}{ll}
\displaystyle A E^{-\Gamma_1}, 
& \displaystyle E<E_{\rm b}, \\[1.5ex]
\displaystyle A E_{\rm b}^{\Gamma_2-\Gamma_1}E^{-\Gamma_2},
& \displaystyle E\ge E_{\rm b},
\end{array}
\right.
\end{equation}
where $\Gamma_1$ and $\Gamma_2$ are the photon indices below and above the break energy $E_{\rm b}$, respectively.

We also use a smoothly broken power-law model (SBPL) for the extra high-energy component,
\begin{equation}
N(E)=A E_{\rm j}^{-\Gamma_1}
\left[\left(\frac{E}{E_{\rm j}}\right)^{\Gamma_1 n}
+\left(\frac{E}{E_{\rm j}}\right)^{\Gamma_2 n}\right]^{-1/n},
\end{equation}
where
\begin{equation}
E_{\rm j}=E_{\rm p,SBPL}
\left(\frac{\Gamma_2-2}{2-\Gamma_1}\right)^{1/[(\Gamma_2-\Gamma_1)n]} .
\end{equation}
Here $E_{\rm p,SBPL}$ is the $\nu F_\nu$ peak energy of the SBPL component. 
The smoothness parameter is fixed at $n=2.69$, following the treatment of GRB 240825A and previous applications to the keV--MeV emission of GRBs \citep{Ryde2010,Zhang2025a}. 
This avoids introducing an additional poorly constrained degree of freedom. 
We therefore use SBPL as a physically motivated representation of a broken extra component.

We employ the Bayesian information criterion \citep{Schwarz1978},
\begin{equation}
{\rm BIC}=-2\ln \mathcal{L}+k\ln N,
\end{equation}
to compare the tested models, where $k$ is the number of free parameters and $N$ is the number of spectral data points. In practice, we use the XSPEC fit statistic in place of $-2\ln \mathcal{L}$.

\begin{figure*}[htbp]
\centering
\begin{minipage}{0.32\textwidth}
  \centering
  \includegraphics[width=\linewidth]{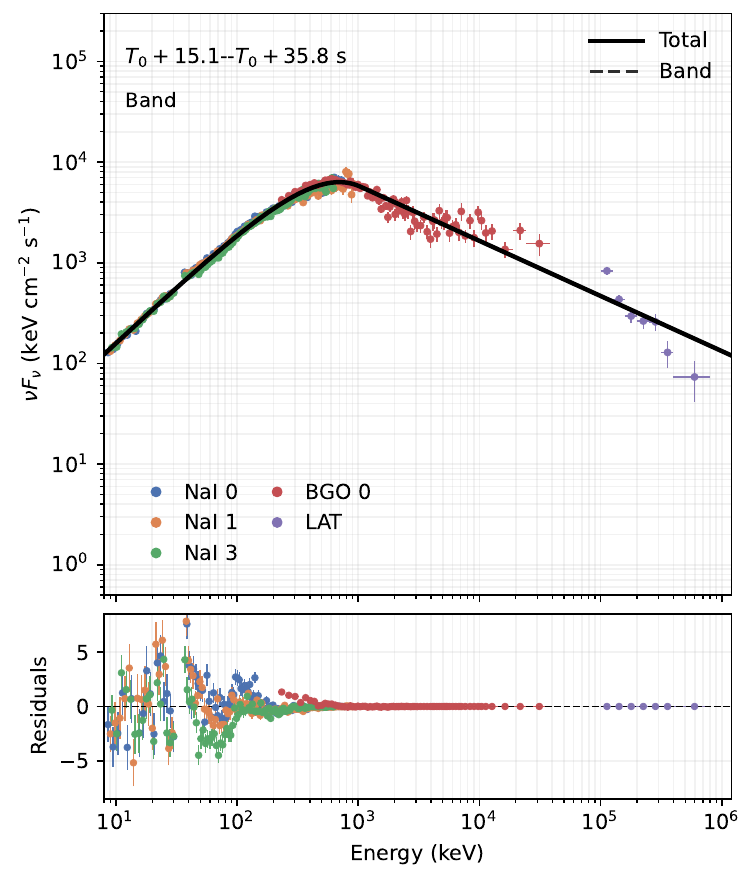}
\end{minipage}\hfill
\begin{minipage}{0.32\textwidth}
  \centering
  \includegraphics[width=\linewidth]{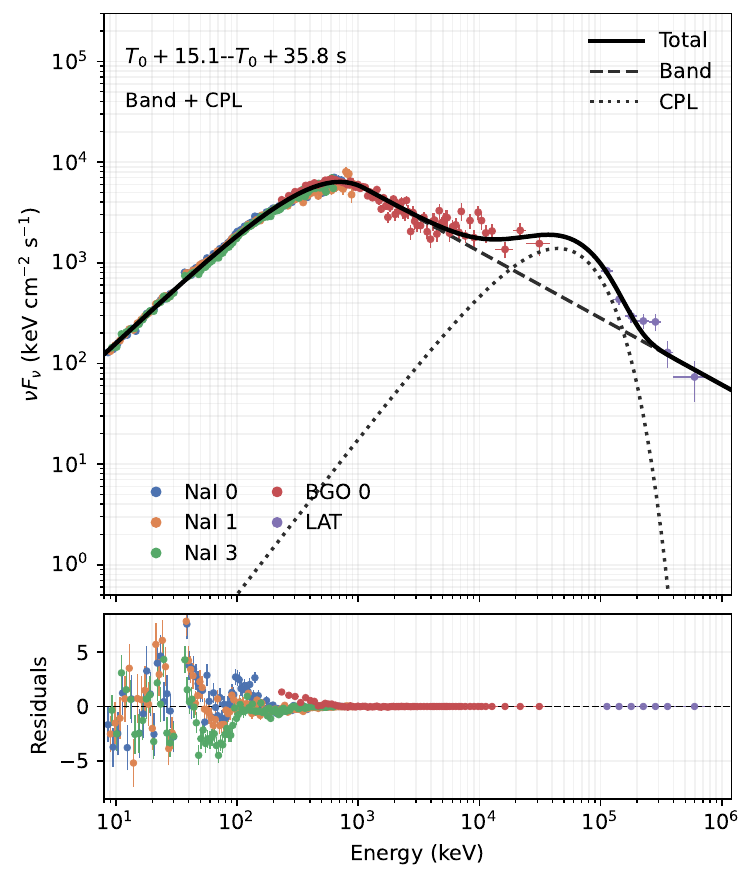}
\end{minipage}\hfill
\begin{minipage}{0.32\textwidth}
  \centering
  \includegraphics[width=\linewidth]{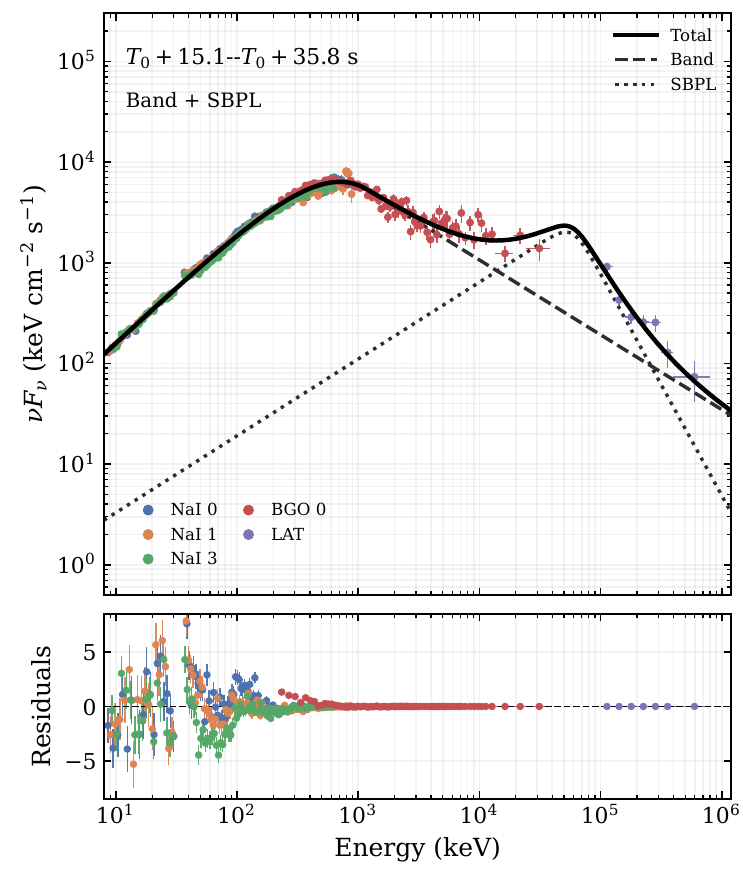}
\end{minipage}
\caption{
Broadband SED of GRB 260226A in the interval
$T_0+15.1$--$T_0+35.8$ s. The three panels show the fits with Band, Band+CPL, and Band+SBPL,respectively. }
\label{fig:sed_15_35}
\end{figure*}

We first focus on the time-integrated spectrum, i.e., the spectrum of the main prompt emission interval $T_0+15.1$--$T_0+35.8$ s (labeled as B--E), which covers the brightest part of the prompt emission.  
The BIC values of the tested models for this time-integrated spectrum are listed in Table~\ref{tab:spec_15_35}, and the corresponding spectra and residuals are shown in Figure~\ref{fig:sed_15_35}.
The single Band model gives a poor fit, with ${\rm BIC}=1661.4$. 
The Band+CPL model gives ${\rm BIC}=1602.8$, while the Band+BPL and Band+SBPL models give ${\rm BIC}=1604.4$ and $1604.6$, respectively. 
Thus, adding an extra high-energy component significantly improves the fit with $\Delta{\rm BIC}\simeq 59$.\footnote{In this work, we adopt $\Delta{\rm BIC}>7$ as the criterion for preferring the model with the lower BIC. 
This threshold corresponds to strong evidence against the model with the higher BIC value \citep{Nunes2017}.}
Meanwhile, the models of CPL, BPL and SBPL  give comparable statistical descriptions for the extra component ($\Delta {\rm BIC} \leq 2$). 
In particular, the BPL and SBPL models give very similar peak energies and spectral slopes. Hereafter, we use SBPL to display the
broken power-law  fit because it provides a smoother representation of the spectral break. 

For the Band+CPL fit, the Band component has $\alpha=-0.87\pm+0.01$, $\beta=-2.66^{+0.04}_{-0.06}$, and $E_{\rm p}= 703.21^{+14.64}_{-14.54}~\rm keV$. 
The additional CPL component has photon index $\Gamma_{\rm cut}=0.45^{+0.71}_{-1.03}$ and cutoff energy $E_{\rm cut}=28.5^{+19.4}_{-10.4}$~MeV. 
The corresponding $\nu F_\nu$ peak of this component is 44.2~MeV, placing the second hump below the GeV band. 
For the Band+SBPL fit, the Band component has $\alpha=-0.87\pm0.01$, $\beta=-2.74\pm0.07$, and $E_{\rm p}=700.57^{+11.94}_{-13.62}$~keV. 
The additional SBPL component has a break energy of $E_{\rm p,SBPL}=51.6^{+13.6}_{-8.2}$ MeV, with photon indices $\Gamma_1=1.24^{+0.13}_{-0.30}$ and $\Gamma_2=4.20^{+1.29}_{-0.46}$. We note that the spectral slope above the break energy in the SBPL model is quite steep, indicating that the spectrum above the break is intrinsically very soft.
The $\nu F_\nu$ peak is located at $\sim 52$ MeV, consistent with the peak energy inferred from the CPL description. 
Assuming $z=1$, the observed 10 keV--1 GeV fluence of the Band+CPL fit is
$7.91\times10^{-4}\ {\rm erg\ cm^{-2}}$, corresponding to
$E_{\rm iso}\simeq2.18\times10^{54}$ erg.
The Band component contributes $E_{\rm iso}^{\rm Band}\simeq1.91\times10^{54}$ erg,
while the additional CPL component contributes
$E_{\rm iso}^{\rm CPL}\simeq2.69\times10^{53}$ erg.

\begin{figure*}[htbp]
\centering
\begin{minipage}{0.24\textwidth}
  \centering
  \includegraphics[width=\linewidth]{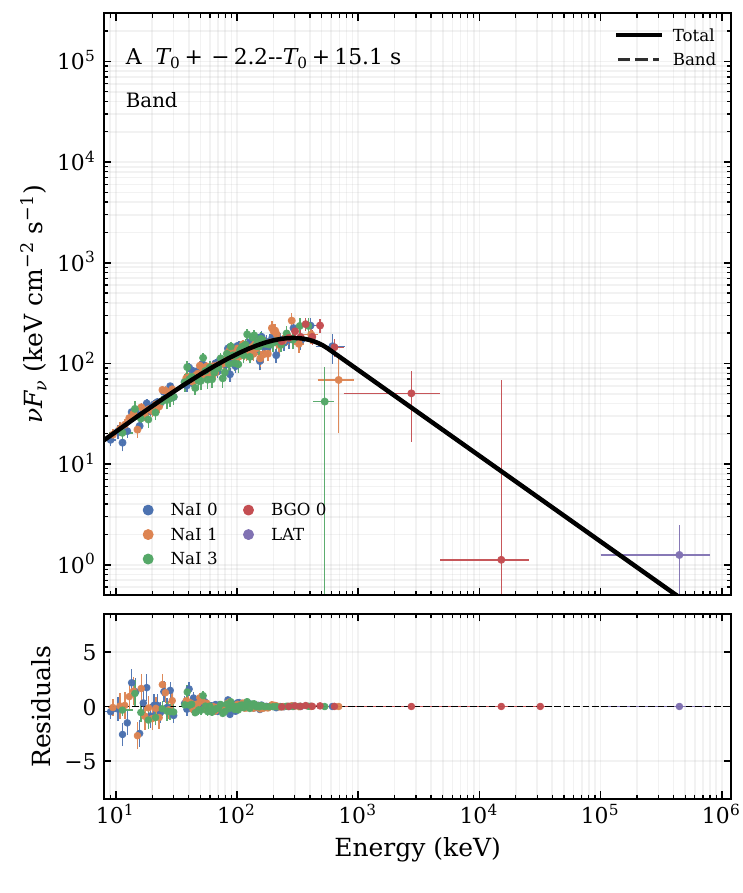}
\end{minipage}
\begin{minipage}{0.24\textwidth}
  \centering
  \includegraphics[width=\linewidth]{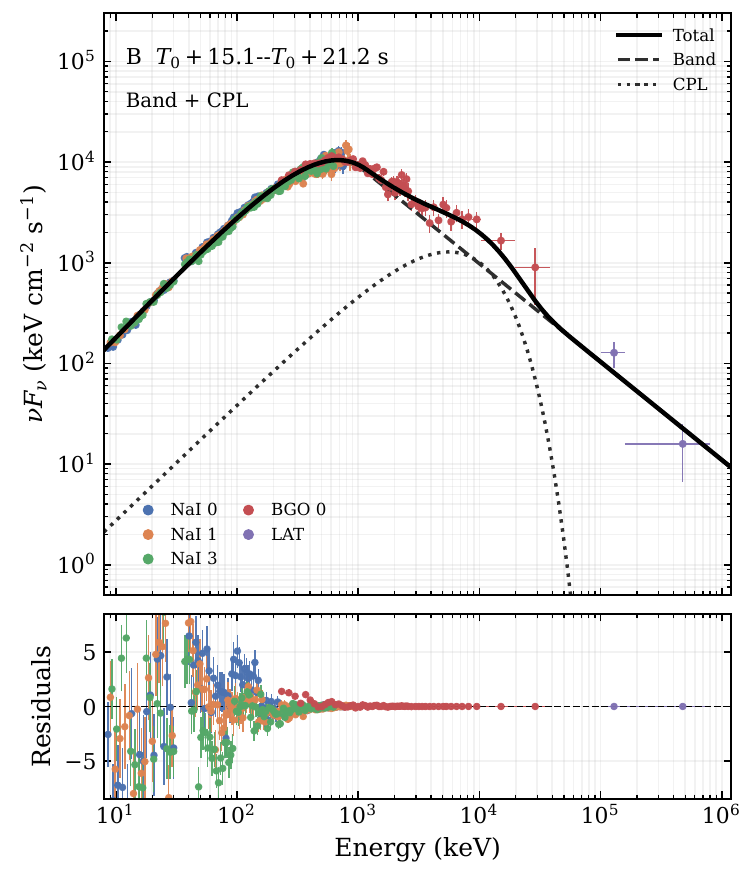}
\end{minipage}
\begin{minipage}{0.24\textwidth}
  \centering
  \includegraphics[width=\linewidth]{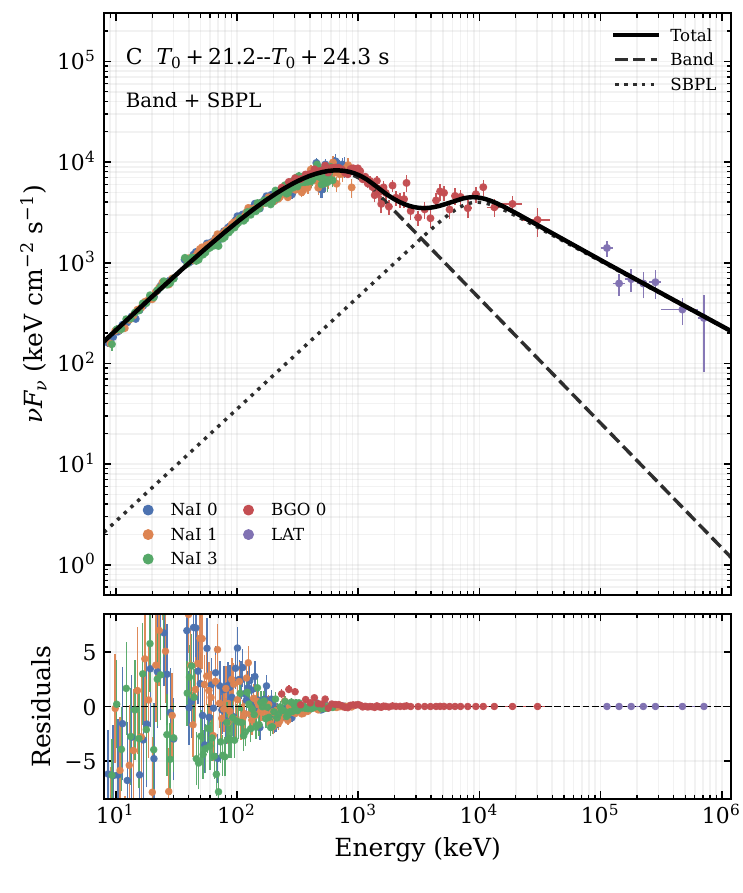}
\end{minipage}
\begin{minipage}{0.24\textwidth}
  \centering
  \includegraphics[width=\linewidth]{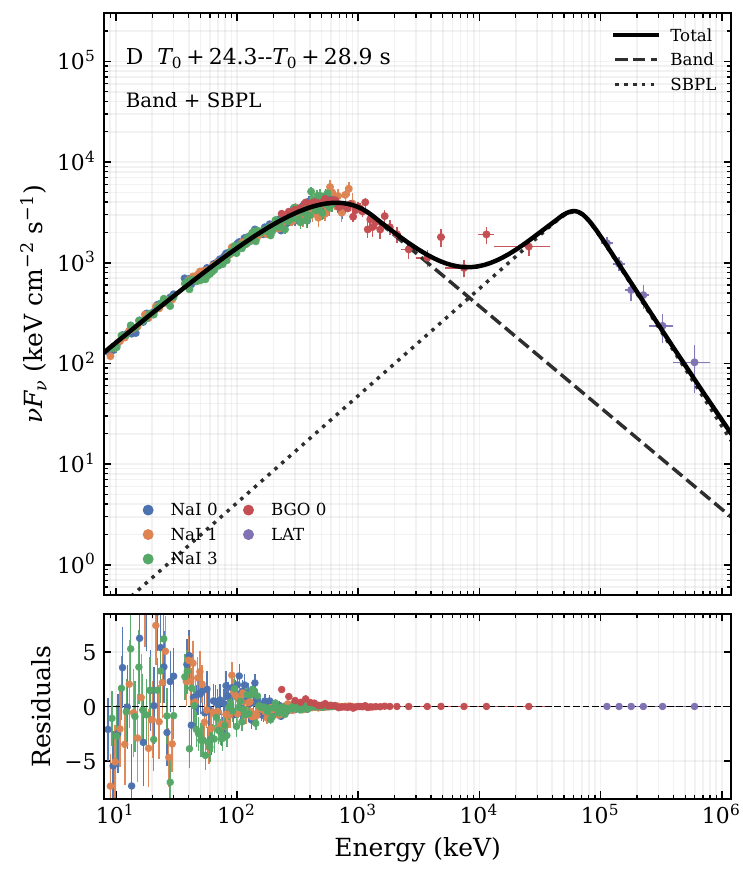}
\end{minipage}

\vspace{2pt}
\begin{minipage}{0.24\textwidth}
  \centering
  \includegraphics[width=\linewidth]{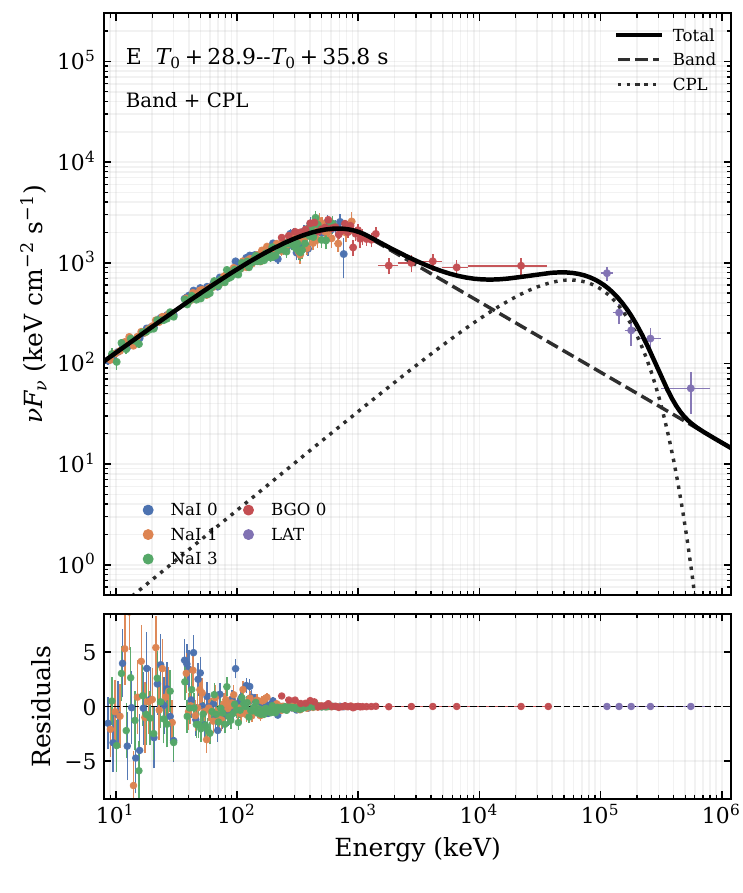}
\end{minipage}
\begin{minipage}{0.24\textwidth}
  \centering
  \includegraphics[width=\linewidth]{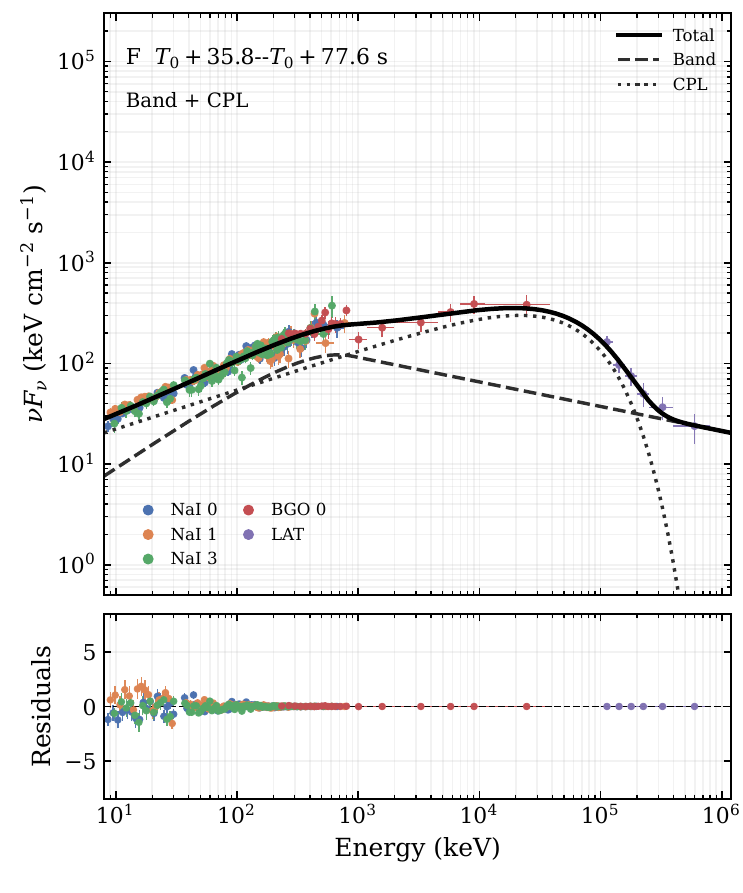}
\end{minipage}
\begin{minipage}{0.24\textwidth}
  \centering
  \includegraphics[width=\linewidth]{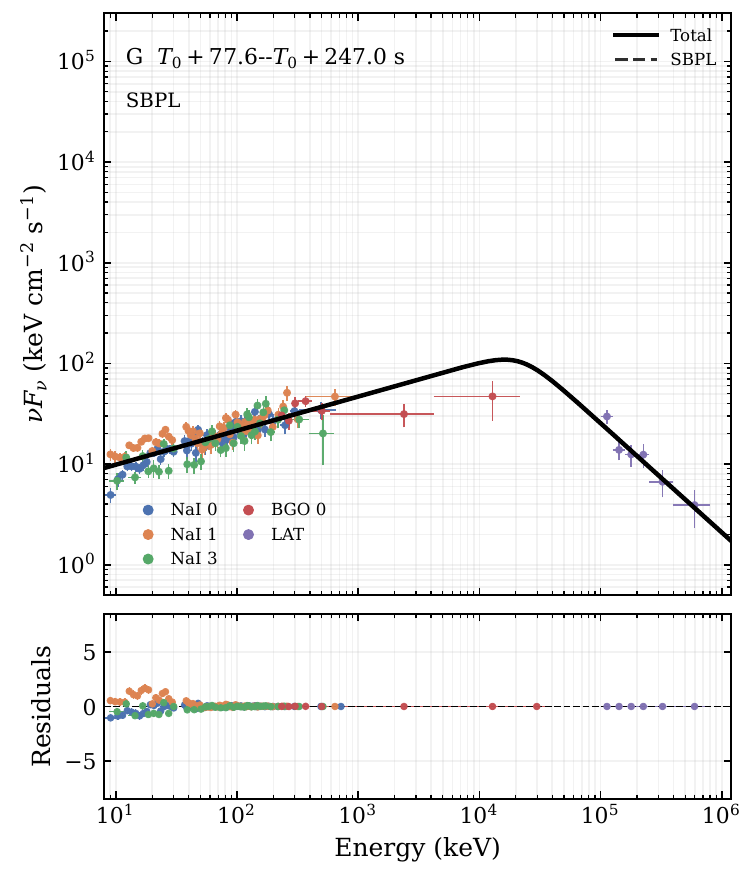}
\end{minipage}
\begin{minipage}{0.24\textwidth}
  \centering
  \includegraphics[width=\linewidth]{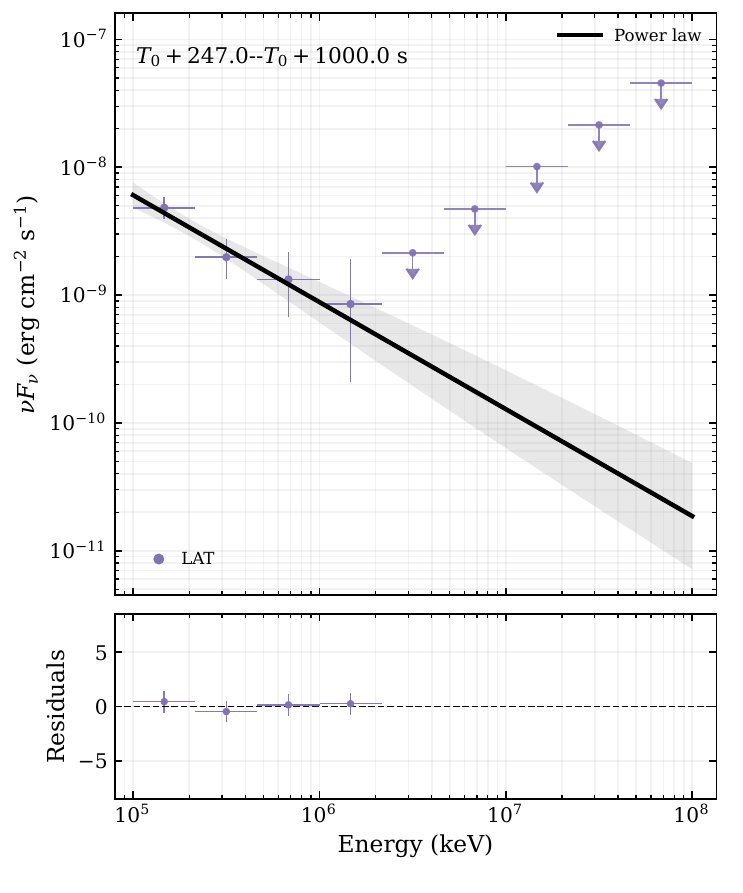}
\end{minipage}

\caption{
Time-resolved broadband SEDs of GRB 260226A for the continuous intervals
from $T_0-2.2$ s to $T_0+247.0$ s, together with the late LAT-only SED for $T_0+247.0$--$T_0+1000$ s in the last panel.
In each panel, the data points are shown together with the best-fit spectral model selected using the BIC criterion. 
Model components are shown with dashed or dotted curves when applicable.}
\label{fig:sed_time_resolved}
\end{figure*}

\begin{figure}[htbp]
\centering
\includegraphics[width=\columnwidth]{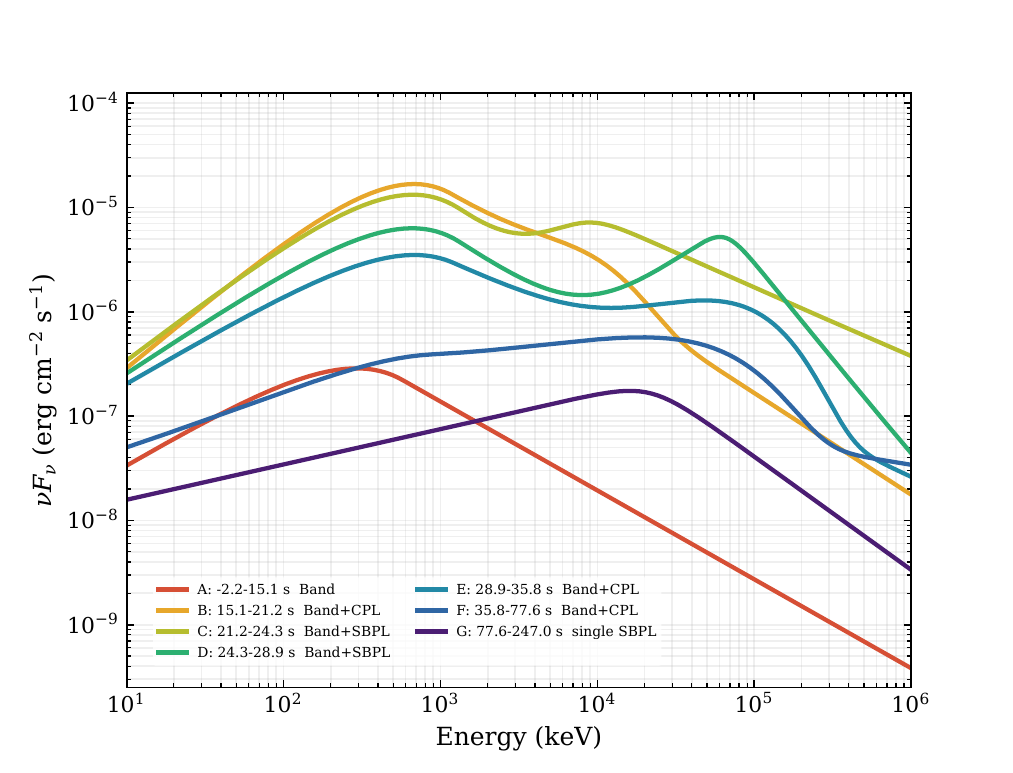}
\caption{
Overlay of the best-fit broadband SED models of GRB 260226A from
$T_0-2.2$ s to $T_0+247.0$ s }
\label{fig:best_seds_overlay}
\end{figure}

Then, we perform time-resolved joint spectral fits to the GBM and LAT data in the seven intervals defined above (Figure~\ref{fig:sed_time_resolved}). 
For interval A, which is before the main pulse, the spectrum is well described by a single Band function.
After the onset of the main pulse, an additional high-energy component is required, and its preferred spectral evolves with time.
At the beginning of the main pulse (interval B), the spectral modelling favors Band+CPL, with $\Delta{\rm BIC}=10.33$ compared with Band+SBPL.
The corresponding $\nu F_\nu$ peak energy of the extra CPL component is $\sim5.6$ MeV.
Around the brightest part of the prompt emission (interval C), the spectrum of the extra component is better described by a SBPL.
Band+SBPL gives the lowest BIC, whereas Band+CPL is disfavored with $\Delta{\rm BIC}=14.27$.
The SBPL peak energy of the extra component is $\sim10$ MeV.
In the later intervals (intervals D--F), Band+CPL and Band+SBPL provide statistically comparable fits, with the peak energy of the extra component increasing to several tens of MeV. For these time intervals, we show the lower-BIC model in Figure~\ref{fig:sed_time_resolved}, although the difference in the BIC value is not large enough to favor one over the other.   We compare CPL and SBPL fits for the extra component for these intervals in the Appendix
Figure~\ref{fig:sed_alternative_models}. Since the difference in the BIC value for the Band+CPL and Band+SBPL models is small, both models provide equally good fit to the data. The spectrum during the interval G is again adequately described by a single Band function, but the peak energy ($E_{\rm p}\simeq 15$ MeV) is considerably higher, making it unlike the low-peak Band component during the earlier intervals. We therefore use a SBPL description for interval G, yielding a low-energy photon index of $\Gamma_1=1.66\pm0.02$, a high-energy photon index of $\Gamma_2=3.09^{+0.43}_{-0.41}$ and peak energy $E_{\rm p,SBPL}=16.08^{+10.47}_{-9.06}$ MeV.
Finally, we analyze the late time LAT-only data from $T_0+247.0$ s to $T_0+1000$ s in the 0.1-100 GeV band.
The spectrum is fitted with a single power-law model, giving a total TS $\simeq 150$ and photon index $\Gamma=2.84\pm0.26$. The BIC values of all the above tested models are summarized in Table~\ref{tab:spec_all}.
An evolution for the spectra during the seven time intervals is shown in Figure~\ref{fig:best_seds_overlay}.

\section{Discussion and Interpretation}
\label{sec:discussion}

\subsection{The extra component}

\begin{figure}[h]
\centering
\includegraphics[width=\columnwidth]{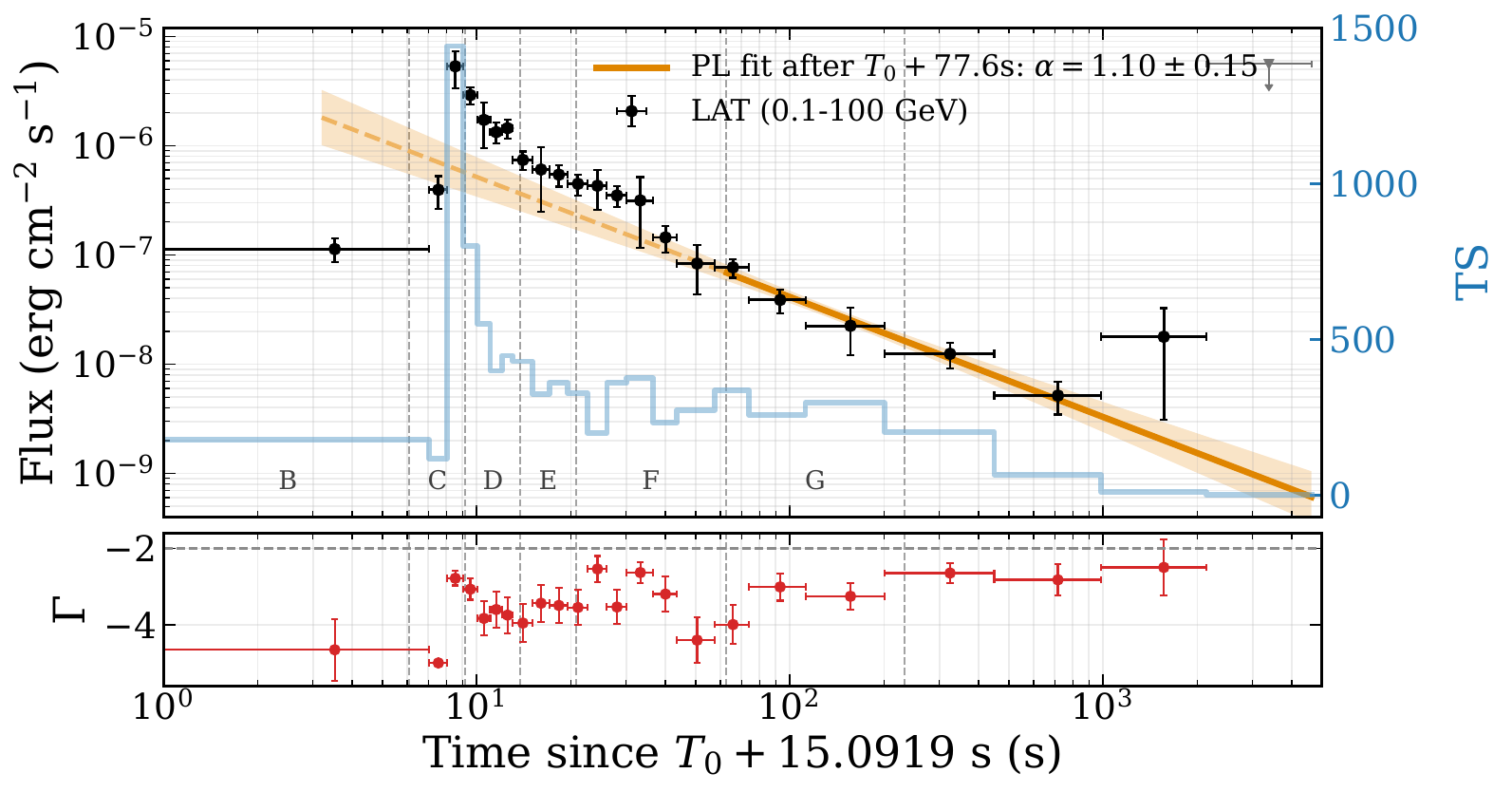}
\caption{LAT light curve of GRB 260226A in the 0.1--100 GeV band.
Top panel: energy fluxes are shown as black points, and 95\% confidence
upper limits are indicated by downward triangles. The blue step curve
shows the TS value in each time bin, with the scale given by the right
axis. The orange solid line represents a single power-law fit to the data at $t-T_0>77.6$ s, using the onset of main pulse as the reference time, where $T_0^{\rm ag}=T_0+15.1$ s. The vertical dashed lines and labels mark intervals B--G.
Bottom panel: photon index evolution for the LAT-detected time bins.
}
\label{fig:lat_lc}
\end{figure}

The origin of the extra spectral component in LAT-detected GRBs remains unknown. 
One possibility is that the delayed extra component is emitted from a forward shock that propagates into the external medium \citep{KumarBarniolDuran2010,Ghisellini2010,Wang2010}, while the Band component arise from internal shocks. 
However, the spectrum of the extra component of GRB 260226A has a break or cutoff at sub-GeV energies, which is inconsistent with the expectation of forward shock emission. 
In addition, the GeV flux during the prompt emission phase is significantly above the extrapolation of the later afterglow emission to earlier times, as shown by Figure~\ref{fig:lat_lc}. 
Therefore, we disfavor this interpretation. 
The other possibility is that the extra component has an internal origin (e.g., internal shock), similar to the Band component. Within this scenario, the high-energy emission could result from SSC emission or external inverse-Compton emission, depending on whether the seed photons for IC scattering are the synchrotron emission of the same population electrons or not. In the latter case, there are two radiation zones: the Band component and the IC emission (the extra component) arise from two different radiation zones. An example for this two-zone scenario is that the Band component is produced by the photosphrere emission and the extra component is produced by IC of the photosphrere emission by relativistic electrons accelerated in internal shocks \citep{Toma2011}. This model has been discussed in the context of GRB090926A \citep{Ackermann2011} and GRB 240825A \citep{Zhang2025a}.

During the period C, the smooth broken power-law model (SBPL) is preferred for the extra component  (see Table~\ref{tab:spec_all}). At later periods of the prompt emission,  the SBPL is as good as CPL model for the extra component, with  slight difference in the BIC values for the fitting (Table~\ref{tab:spec_all}). 
If the Band component is interpreted as synchrotron emission and the extra component as its SSC counterpart, the two peak frequencies should satisfy $\gamma_m\simeq\sqrt{\nu_{\rm SSC}/\nu_{\rm syn}}$, where $\gamma_m$ is the minimum Lorentz factor of the injected electrons.
For GRB 260226A, the time-integrated main-pulse fit gives $h\nu_{\rm syn}\simeq 622$ keV and $h\nu_{\rm extra}\simeq 44$ MeV, corresponding to $\gamma_m\simeq 8$. 
Such a low $\gamma_m$ would require an extremely large magnetic field, and is therefore difficult to explain with a standard optically thin synchrotron+SSC interpretation. Thus, GRB 260226A may be understood in a picture similar to that proposed for GRB 240825A: the extra component is unlikely to be the ordinary SSC counterpart of the Band component, but 
may instead be associated with some other process\citep{Zhang2025a}.
One scenario for this process is the inverse-Compton scattering of photospheric radiation.
In the photospheric-emission scenario of the Band component \citep[e.g.,][]{MeszarosRees2000,Ryde2010},  the extra component could be produced by inverse-Compton scattering of photospheric photons by relativistic electrons accelerated at  larger radii, e.g., in an internal-shock or dissipation region above the photosphere \citep{Beloborodov2010,Gao2009,Ryde2010,Toma2011}. In the internal shock scenario, $\gamma_m\simeq \epsilon_e \frac{p-2}{p-1}\frac{m_p}{m_e} (\Gamma_{\rm in}-1)$, where $\Gamma_{\rm in}$ is the Lorentz factor of the internal shock. For typical value of $\epsilon_e=0.1$, $p=2.2-2.5$, we need $\Gamma_{\rm in}-1\sim 0.1-0.4$, implying that the internal shock should be at most mildly-relativistic for the case of GRB 260226A. 

During the period B, the extra component is best described by a CPL, with a cutoff energy of $\sim 5~{\rm MeV}$ (see Table~\ref{tab:spec_all}). Interpreting this cutoff as the $\gamma\gamma$ absorption, the low cutoff energy implies a Lorentz factor \citep{Lithwick2001, Li-Zhuo2010}
\begin{equation}
\Gamma\simeq \frac{E_c}{m_e c^2}=10.
\end{equation}
For such low $E_c$ pulse, the energy of target photons for $\gamma\gamma$ absorption is comparable to $E_c$ (i.e., $E_c\ga \Gamma^2 m_e^2 c^4/E_c$).
The lower Lorentz factor may explain the soft spectrum at early time and delayed onset of the GeV photons. In the photosphere IC model for the extra component, the Band component is produced by the photosphere emission and the photosphere emission region could have a higher Lorentz factor than the IC emission region, so the Band component extends to a few 100 MeV. During the late time intervals, significant GeV emission is detected  in the extra component and the Lorentz factor should be $\Gamma\ga 100$. This may reflect that the  initial pulse results from the  jet breakout out of the star, which causes  more baryon loading entrainment (and hence a lower Lorentz factor), while at later times, the jet is more clean and has a higher Lorentz factor. 


\subsection{LAT Extended GeV Emission}
For the long-term LAT emission, we perform an unbinned likelihood analysis of the LAT ($>100$ MeV) data. The LAT light curve is shown in Figure~\ref{fig:lat_lc}.
Only bins with ${\rm TS}>9$ were retained for temporal fitting. In the LAT light curve, the emission is initially weak, then rises rapidly and followed by a smooth, long-lasting decay. 
The light curve decays as a single power law, indicating an afterglow origin for the extended GeV emission.
As shown by the GBM count-rate light curve, the burst has an early quiescent phase before the onset of the main pulse. 
We therefore use this onset time of the main pulse, $T_0^{ag} = T_0+15.1$ s, as the reference time for the afterglow emission modelling \citep{Kobayashi2007}.
We fitted the LAT data points starting from $t-T_0=77.6$ s, corresponding to the end of interval F  of the prompt emission, with a single power-law function,
\begin{equation}
    F(t) \propto (t-T_0^{ag})^{-\alpha}.
\end{equation}
The best-fit decay index is $\alpha=1.10\pm0.15$.
The late-time LAT light curve is well described by a single power-law decay, suggesting that the GeV emission after $T_0+77.6$~s is dominated by the external-shock afterglow component. The GeV flux at earlier time exceeds the extrapolation of the power-law behavior, indicating an internal origin for the GeV emission at such early times.

\section{Conclusions}
\label{sec:conclusion}

We have performed a  joint GBM--LAT spectral analysis of GRB 260226A. During the main prompt emission phase, both the time-integrated spectrum and  time-resolved spectrum show an extra hard component 
beyond the Band function component. During the period B, the extra component is best describe by a CPL model, while during the period C, the SBPL is preferred. For the time-integrated spectrum, Band+CPL and Band+SBPL provide statistically similar descriptions.  In either case, the peak energy in the extra component is about 50 MeV. Thus, GRB 260226A adds a new example, after GRB 190114C and GRB 240825A, of GRBs that have an extra component peaking at sub-GeV band. 

The low ratio between the peak energy of extra component and the Band component is difficult to explain with a standard one-zone synchrotron+SSC picture, and may instead point to two-zone scenario, where the Band component and the extra component are produced at different emission regions. One possible scenario is that the Band component is produced by the photosphere emission and the extra component is produced by the IC emission of relativistic electrons accelerate at internal shocks outside the photosphere\citep{Toma2011}. 

\begin{acknowledgments}
This work is supported by the
National Natural Science Foundation of China (grant
Nos.  12121003 and 12333006) and  the
Fundamental Research Funds for the Central Universities (KG202502). We are grateful to the High Performance Computing Center of
Nanjing University for doing the numerical calculations in this
paper on its blade cluster system.
\end{acknowledgments}

\appendix
\section{Supplementary Time-resolved Spectral Fits}
\label{app:time_resolved_seds}

\begin{figure*}[t]
\centering
\begin{minipage}{0.32\textwidth}
  \centering
  (a) Interval D\par\smallskip
  \includegraphics[width=\linewidth]{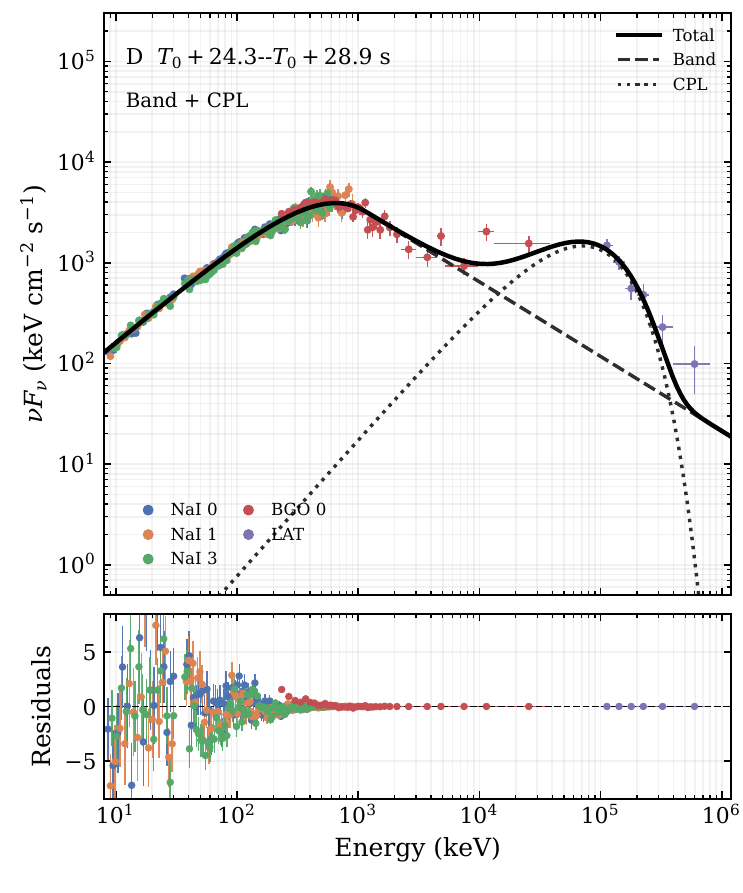}\\[-0.2em]
  \includegraphics[width=\linewidth]{SBPL_main/D_24.3_28.9_best_sed.pdf}
\end{minipage}\hfill
\begin{minipage}{0.32\textwidth}
  \centering
  (b) Interval E\par\smallskip
  \includegraphics[width=\linewidth]{best_seds/E_28.9_35.8_best_sed.pdf}\\[-0.2em]
  \includegraphics[width=\linewidth]{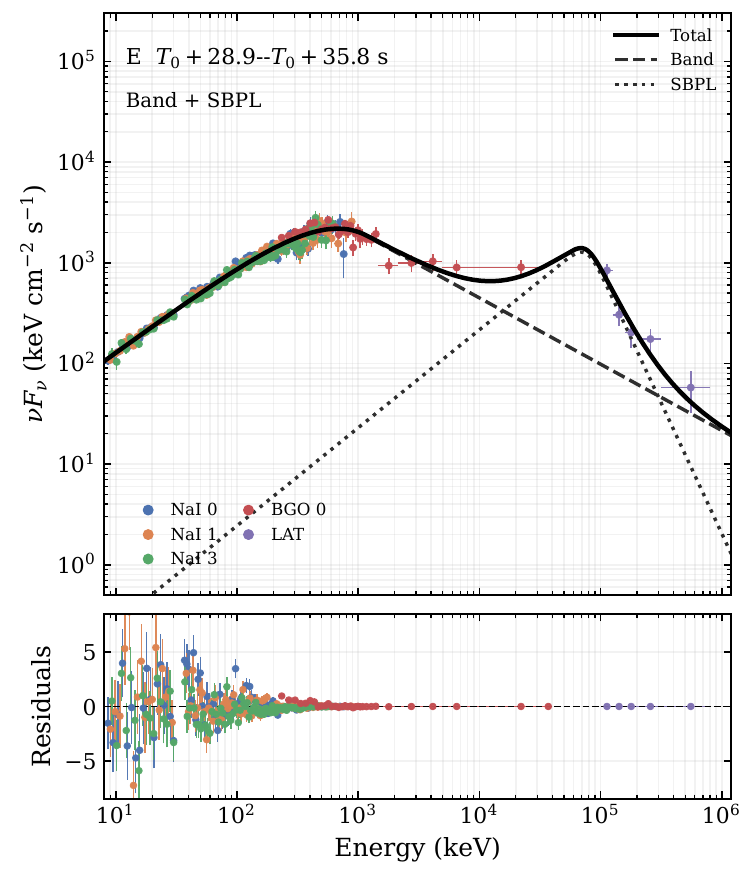}
\end{minipage}\hfill
\begin{minipage}{0.32\textwidth}
  \centering
  (c) Interval F\par\smallskip
  \includegraphics[width=\linewidth]{best_seds/F_35.8_77.6_best_sed.pdf}\\[-0.2em]
  \includegraphics[width=\linewidth]{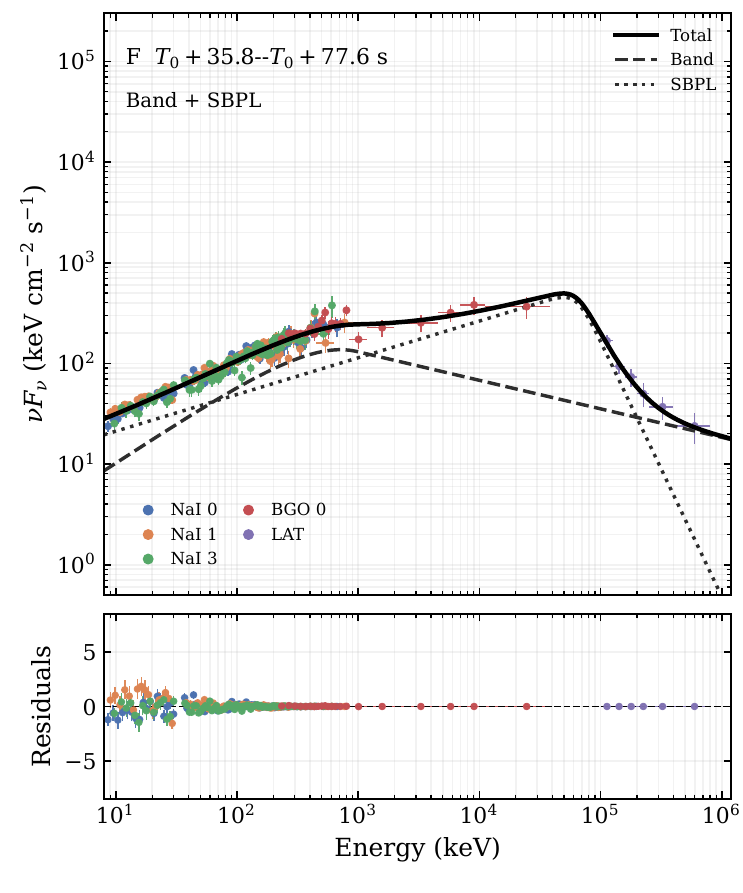}
\end{minipage}

\caption{
Comparison between Band+CPL and Band+SBPL fits for selected
time-resolved intervals. Panels a, b, and c show intervals D, E, and F, respectively.
}
\label{fig:sed_alternative_models}
\end{figure*}

In this Appendix, we present the complete time-resolved spectral fitting results. 
The main text adopts the lowest-BIC model among all tested models in each interval.
However, in some intervals the two-component models have comparable BIC values and provide similar descriptions of the data.
For the Band+CPL fits, the $\nu F_\nu$ peak energies of the CPL component, calculated from the best-fit model spectra, are $ 70.97$, $56.12$, and $21.60$ MeV for intervals D, E, and F, respectively.
For the Band+SBPL fits, the corresponding $\nu F_\nu$ peak energies of the SBPL component are $60.54$, $69.58$, and $49.77$ MeV, respectively.
The $\Delta{\rm BIC}$ values of the Band+SBPL fits relative to the Band+CPL models are $0.12$, $1.99$, and $5.04$ for intervals D, E, and F, respectively.
We therefore show representative alternative fits in
Figure~\ref{fig:sed_alternative_models} to illustrate that the two models give similar two-component descriptions with comparable fit statistics.


These supplementary figures are not used to redefine the fiducial time-resolved spectral sequence, but provide a visual comparison between models with similar statistical performance.


\bibliography{sample701}{}
\bibliographystyle{aasjournalv7}

\begin{table*}[t]
\centering
\scriptsize
\setlength{\tabcolsep}{3pt}
\caption{Spectral fitting results for the time-integrated spectrum during the interval $T_0+15.1$--$T_0+35.8$ s.}
\label{tab:spec_15_35}
\begin{tabular}{lcccc}
\hline
\hline
Model & Band & Band+CPL & Band+BPL &  Band+SBPL\\
\hline
Band function & & & &\\
$\alpha$
& $-0.87\pm0.01$
& $-0.87\pm0.01$
& $-0.87^{+0.02}_{-0.01}$ 
& $-0.87\pm0.01$\\
$\beta$
& $-2.55\pm0.02$
& $-2.66^{+0.04}_{-0.06}$
& $-2.74^{+0.12}_{-0.11}$ 
& $-2.74\pm0.07$\\
$E_{\rm p}$ (keV)
& $699.39^{+13.79}_{-13.22}$
& $703.21^{+14.64}_{-14.54}$
& $700.65^{+19.75}_{-22.44}$
& $700.57^{+11.94}_{-13.62}$\\
\hline
CPL & & & \\
$\Gamma_{\rm cut}$
& --
& $0.45^{+0.71}_{-1.03}$
& --
& -- \\
$E_{\rm cut}$ (MeV)
& --
& $28.5^{+19.4}_{-10.4}$
& -- 
& -- \\
\hline
BPL & & & \\
$\Gamma_1$
& --
& --
& $1.23^{+0.17}_{-0.14}$ 
& -- \\
$E_{\rm br}$ (MeV)
& --
& --
& $59.3^{+30.9}_{-14.3}$ 
& -- \\
$\Gamma_2$
& --
& --
& $4.22$ 
& -- \\
\hline
SBPL & & & & \\
$\Gamma_1$
& --
& --
& --
& $1.24^{+0.13}_{-0.30}$ \\
$E_{\rm p,SBPL}$ (MeV)
& --
& --
& --
& $51.60^{+13.57}_{-8.23}$ \\
$\Gamma_2$
& --
& --
& --
& $4.20^{+1.29}_{-0.46}$ \\
$n$
& --
& --
& --
& $2.69$ (fixed) \\
\hline
Statistic/dof
& $1636.88/459$
& $1559.79/456$
& $1555.30/455$
& $1555.50/455$ \\
BIC
& $1661.43$
& $1602.75$
& $1604.40$
& $1604.60$ \\
$\Delta$BIC
& $58.68$
& $0.00$
& $1.65$
& $1.85$ \\
\hline
\end{tabular}%

\vspace{2pt}
\end{table*}

\begin{sidewaystable*}[p]
\centering
\footnotesize
\setlength{\tabcolsep}{4pt}
\renewcommand{\arraystretch}{1.1}
\caption{Spectral fitting results for the main intervals A--G. 
Uncertainties are quoted when constrained by XSPEC. 
For each interval, $\Delta{\rm BIC}$ is calculated relative to the minimum BIC among the listed models. For interval G, the CPL, BPL, and SBPL entries are single-component fits.}
\label{tab:spec_all}
\resizebox{1\textwidth}{!}{%
\begin{tabular}{llccccccc}
\hline
\hline
Model & Par. & A & B & C & D & E & F & G \\
 & s & -2.2--15.1 & 15.1--21.2 & 21.2--24.3 & 24.3--28.9 & 28.9--35.8 & 35.8--77.6 & 77.6--247.0 \\
 & preferred & Band & Band+CPL & Band+BPL(SBPL) & Band+BPL(SBPL) & Band+CPL & Band+CPL & Band \\
\hline
Band & $\alpha$ & $-1.11\pm0.05$ & $-0.74\pm0.01$ & $-0.85\pm0.01$ & $-0.99\pm0.01$ & $-1.12\pm0.01$ & $-1.47$ & $-1.65\pm0.02$ \\
 & $\beta$ & $-2.85^{+0.23}_{-0.41}$ & $-2.84\pm0.05$ & $-2.44\pm0.03$ & $-2.37\pm0.03$ & $-2.40^{+0.03}_{-0.04}$ & $-2.41^{+0.01}_{-0.21}$ & $-3.15^{+0.41}_{-0.43}$ \\
 & $E_{\rm p}$ (keV) & $291.85^{+59.67}_{-47.69}$ & $683.27^{+16.19}_{-16.12}$ & $646.59^{+25.04}_{-23.13}$ & $629.00^{+31.41}_{-29.62}$ & $665.50^{+50.58}_{-45.43}$ & $2733.88$ & $14521.34^{+4214.20}_{-5988.89}$ \\
 & stat & $310.67/459$ & $1076.76/459$ & $737.67/459$ & $743.39/459$ & $560.70/459$ & $385.06/459$ & $553.85/459$ \\
 & BIC & $335.22$ & $1101.31$ & $762.23$ & $767.94$ & $585.25$ & $409.61$ & $578.40$ \\
 & $\Delta$BIC & $0.00$ & $8.27$ & $15.92$ & $66.83$ & $10.66$ & $14.53$ & $0.00$ \\
\hline
Band+CPL & $\alpha$ & $5.00$ & $-0.74\pm0.01$ & $-0.85\pm0.01$ & $-1.00\pm0.01$ & $-1.12^{+0.03}_{-0.01}$ & $-1.20^{+0.41}_{-0.28}$ & -- \\
 & $\beta$ & $-2.78^{+0.28}_{-0.41}$ & $-2.97^{+0.07}_{-0.05}$ & $-2.51^{+0.05}_{-0.06}$ & $-2.74^{+0.09}_{-0.06}$ & $-2.70$ & $-2.24^{+0.11}_{-0.03}$ & -- \\
 & $E_{\rm p}$ (keV) & $460.28$ & $662.43^{+21.56}_{-21.22}$ & $644.20^{+24.15}_{-25.64}$ & $651.36^{+31.84}_{-30.25}$ & $672.92^{+55.36}_{-62.34}$ & $683.67^{+884.44}_{-472.31}$ & -- \\
 & $\Gamma_{\rm cut}$ & $0.99^{+0.11}_{-0.12}$ & $0.85^{+0.22}_{-0.50}$ & $0.55^{+0.60}_{-0.57}$ & $0.64^{+0.55}_{-0.82}$ & $1.01^{+0.37}_{-0.47}$ & $1.61^{+0.06}_{-0.20}$ & $1.69\pm0.02$ \\
 & $E_{\rm cut}$ (MeV) & $0.18^{+0.10}_{-0.06}$ & $4.85^{+1.60}_{-0.59}$ & $24.79^{+16.08}_{-7.65}$ & $52.29^{+10.71}_{-10.66}$ & $56.82$ & $55.45^{+7.54}_{-13.82}$ & $69.91^{+7.72}_{-6.54}$ \\
 & stat & $304.68/456$ & $1050.08/456$ & $717.61/456$ & $659.66/456$ & $531.62/456$ & $352.12/456$ & $605.73/460$ \\
 & BIC & $347.65$ & $1093.05$ & $760.58$ & $702.63$ & $574.59$ & $395.08$ & $624.14$ \\
 & $\Delta$BIC & $12.43$ & $0.00$ & $14.27$ & $1.52$ & $0.00$ & $0.00$ & $45.74$ \\
\hline
Band+BPL & $\alpha$ & $-1.11\pm0.05$ & $-0.74\pm0.01$ & $-0.86^{+0.02}_{-0.01}$ & $-1.00\pm0.01$ & $-1.09^{+0.07}_{-0.05}$ & $-1.21^{+0.32}_{-0.04}$ & -- \\
 & $\beta$ & $-2.85^{+0.23}_{-0.41}$ & $-2.97^{+0.09}_{-0.14}$ & $-3.08^{+0.06}_{-0.09}$ &
 $-3.01^{+0.29}_{-1.36}$ & $-2.94$ & $-2.28$ & -- \\
 & $E_{\rm p}$ (keV) & $291.75^{+59.75}_{-47.54}$ & $673.49^{+18.70}_{-17.78}$ & $645.09^{+30.11}_{-29.27}$ & $655.26^{+32.70}_{-30.53}$ & $650.90^{+101.44}_{-93.04}$ & $718.29^{+7140.86}_{-383.82}$ & -- \\
 & $\Gamma_1$ & $8.92$ & $0.98^{+0.22}_{-0.46}$ & $0.97^{+0.23}_{-0.32}$ & $0.93^{+0.60}_{-0.58}$ & $1.52^{+0.10}_{-0.43}$ & $1.63^{+0.06}_{-0.06}$ & $1.66\pm0.02$ \\
 & $E_{\rm br}$ (MeV) & $683.30$ & $7.18$ & $9.83^{+2.18}_{-1.89}$ & $64.97^{+35.82}_{-15.74}$ & $78.17$ & $63.41^{+25.82}_{-33.97}$ & $21.96^{+14.09}_{-12.48}$ \\
 & $\Gamma_2$ & $-1.30$ & $3.71$ & $2.64^{+0.10}_{-0.12}$ & $3.88^{+0.56}_{-0.43}$ & $3.92^{+0.66}_{-0.51}$ & $4.52^{+3.46}_{-1.41}$ & $3.09^{+0.43}_{-0.40}$ \\
 & stat & $310.66/455$ & $1054.92/455$ & $698.05/455$ & $652.01/455$ & $527.26/455$ & $350.99/455$ & $556.71/459$ \\
 & BIC & $359.76$ & $1104.02$ & $747.15$ & $701.11$ & $576.36$ & $400.09$ & $581.26$ \\
 & $\Delta$BIC & $24.55$ & $10.97$ & $0.84$ & $0.00$ & $1.78$ & $5.01$ & $2.86$ \\
\hline
Band+SBPL & $\alpha$ & $-1.11^{+0.02}_{-0.04}$ & $-0.74\pm0.01$ & $-0.86\pm0.01$ & $-1.00\pm0.01$ & $-1.12^{+0.08}_{-0.01}$ & $-1.21^{+0.18}_{-0.19}$ & -- \\
& $\beta$ & $-3.85$ & $-2.98^{+0.09}_{-0.19}$ & $-3.24^{+0.31}_{-0.68}$ & $-3.01^{+0.22}_{-0.56}$ & $-2.66^{+0.13}_{-0.79}$ & $-2.28^{+0.07}_{-0.25}$ & -- \\
& $E_{\rm p}$ (keV) & $299.76^{+32.82}_{-31.85}$ & $673.43^{+18.16}_{-19.48}$ & $643.62^{+19.93}_{-20.77}$ & $655.27^{+19.71}_{-20.20}$ & $672.59^{+66.43}_{-28.10}$ & $715.99^{+633.85}_{-306.82}$ & -- \\
& $\Gamma_1$ & $0.61$ & $0.97^{+0.22}_{-0.27}$ & $0.88^{+0.20}_{-0.45}$ & $0.94^{+0.52}_{-0.55}$ & $1.03^{+0.49}_{-0.06}$ & $1.63^{+0.04}_{-0.10}$ & $1.66\pm0.02$ \\
& $E_{\rm p,SBPL}$ (MeV) & $681.19$ & $7.06$ & $9.64^{+1.92}_{-1.83}$ & $60.54^{+12.90}_{-8.13}$ & $69.58^{+22.82}_{-18.18}$ & $49.77^{+16.96}_{-18.34}$ & $16.08^{+10.47}_{-9.06}$ \\
& $\Gamma_2$ & $2.29$ & $3.77$ & $2.65^{+0.08}_{-0.07}$ & $3.88^{+0.31}_{-0.27}$ & $4.59$ & $4.54^{+1.76}_{-1.14}$ & $3.09^{+0.43}_{-0.41}$ \\
& $n$ (fixed) & $2.69$ & $2.69$ & $2.69$ & $2.69$ & $2.69$ & $2.69$ & $2.69$ \\
& stat & $309.93/455$ & $1054.28/455$ & $697.21/455$ & $652.13/455$ & $527.48/455$ & $351.02/455$ & $556.12/459$ \\
& BIC & $359.03$ & $1103.38$ & $746.31$ & $701.23$ & $576.58$ & $400.12$ & $580.67$ \\
& $\Delta$BIC & $23.81$ & $10.33$ & $0.00$ & $0.12$ & $1.99$ & $5.04$ & $2.28$ \\
\hline
\end{tabular}%
}
\vspace{2pt}
\parbox{\textwidth}{\footnotesize
$^{\dagger}$ Uncertainties are not reported when the parameter is not well constrained.
The SBPL smoothness parameter $n$ was fixed at 2.69.
}
\end{sidewaystable*}




\end{document}